\documentclass[prb,aps,superscriptaddress,twocolumn]{revtex4}%
\usepackage{graphicx}
\usepackage{dcolumn}
\usepackage{bm}
\usepackage{color}
\usepackage{ulem}
\usepackage{amsmath}
\usepackage{amsfonts}
\usepackage{amssymb}%
\providecommand{\U}[1]{\protect\rule{.1in}{.1in}}

\begin{document}
\title{Lifting mean-field degeneracies in anisotropic  classical spin systems}
\author{Yuriy Sizyuk}
\affiliation{School of Physics and Astronomy, University of Minnesota, Minneapolis, MN
55116, USA}
\author{Natalia B. Perkins}
\affiliation{School of Physics and Astronomy, University of Minnesota, Minneapolis, MN
55116, USA}
\author{Peter W\"{o}lfle}
\affiliation{Institute for Condensed Matter Theory and Institute for Nanotechnology,
Karlsruhe Institute of Technology, D-76128 Karlsruhe, Germany}

\begin{abstract}
 In this work, we propose a method for calculating the free energy of anisotropic classical spin systems. We use a Hubbard-Stratonovich transformation to express the
partition function of a generic bilinear super-exchange Hamiltonian in terms of a functional integral
over classical time-independent fields. As an example, we consider an anisotropic spin-exchange
Hamiltonian on the cubic lattice as is found for compounds with strongly correlated electrons in
multiorbital bands and subject to strong spin-orbit interaction. We calculate the contribution of
Gaussian  spin fluctuations to the free energy. While the mean-field solution
of ordered states for such systems usually has full rotational symmetry, we show here that the
fluctuations lead to a pinning of the spontaneous magnetization along some preferred direction of
the lattice.

\end{abstract}
\maketitle

\section{Introduction}

Recent research activities on transition metal oxides suggest that the
interplay of the strong spin-orbit coupling (SOC), crystal field (CF)
interactions, and electron correlations may lead to compasslike anisotropic
interactions between magnetic degrees of freedom.\cite{nussinov15} These
anisotropic interactions have a generic form $J_{ij}^{\alpha}S_{i}^{\alpha
}S_{j}^{\alpha}$ in which $\alpha$ depends on the direction of the particular
link or bond and $S$ denotes spin or pseudospin degrees of freedom describing magnetic or orbital
degrees of freedom.

The models in which compasslike anisotropies are dominating, or also the pure
compass models, have been known for a long time. These models appear naturally
in strongly correlated electron systems as minimal models to account for
interactions between pseudospins describing orbital degrees of
freedom.\cite{kk82,nussinov04,nussinov05,cincio10,nasu12,oles13}
The compass-like anisotropies also arise as interactions between magnetic
degrees of freedom in systems with strong SOC, which might be realized in 4d
and 5d transition metal oxides.\cite{jackeli09} However, in these systems, due
to the extended nature of 4d and 5d orbitals, the compass interactions are
always accompanied by the usual SU(2) symmetric Heisenberg-type exchange.
These models are especially interesting because while the pure compasslike
models are rare, the combined Heisenberg-compass models have been shown to be
minimal models describing the magnetic properties of various materials. A
review of the different realizations of compass models,
\cite{kk82,nussinov04,nussinov05,cincio10,nasu12,oles13,jackeli09,batista05,biskup05,kitaev06,jackeli10,chern10,
oitmaa10,fabien10,khali2001,wenzel10, perkins14,sizyuk14,plakida14,gerlach15}
their physical motivations, symmetries, unconventional orderings and
excitations may be found in the recent paper by Nussinov and van den
Brink.\cite{nussinov15}

One of the common features induced by compasslike anisotropies is
frustration, arising from a competition of interactions along different
directions and leading to the macroscopic degeneracy of the classical ground
state and in addition to rich quantum behavior. In many cases, the pure
compass models do not show conventional magnetic ordering because the
degeneracy of the classical ground state is connected to discrete sliding
symmetries of the model.\cite{nussinov05,batista05} Because these symmetries
are intrinsic symmetries of the model, they can not be lifted by the order-by
disorder mechanisms. Instead, the direct consequence of the existence of these
symmetries is that the natural order parameters for pure compass models are
nematic, which are invariant under discrete sliding symmetries.

The nematic order present in the compass model is fragile and is easily
destroyed by the presence of the isotropic Heisenberg interaction which breaks
some of the intrinsic symmetries of the model. In Heisenberg-compass models,
some of the degeneracies become accidental. In these models, the true magnetic
order might be selected by fluctuations via an order by disorder mechanism,
removing accidental degeneracies and determining both the nature and the
direction of the order parameter. Despite the simplicity of these models, the
interplay of the Heisenberg and compass interaction leads to very rich phase
diagrams even in the simplest case of the square lattice.\cite{fabien10} For
classical systems this mechanism requires finite temperatures, where entropic
contributions of fluctuations to the free energy become effective.

In this work, we will be interested in studying the directional ordering
transitions in the Heisenberg-compass model on the cubic
lattice.\cite{khali2001} From a historical perspective, the three-dimensional
90$^{\circ}$-compass model was the first model of this kind proposed by Kugel
and Khomskii\cite{kk82} in the context of the ordering of the $t_{2g}$
orbitals in transition metal oxides with perovskite structure and then studied
in more details by Khaliullin\cite{khali2001} in application to LaTiO$_{3}$.
The formal procedure which we will be using here is based on the derivation of
the fluctuational part of the free energy by integrating out the Gaussian
fluctuations, and determining which orientations of the vector order parameter
correspond to the free energy minimum. To do so, we first express the
partition function as a functional integral over classical fields. In this first paper,
we consider classical spins at finite temperature. Our
starting point in evaluating this exact representation of the partition
function is the mean-field solution, which usually does not reflect the
anisotropic character of the interaction referring to the crystal lattice
axes. As a next step, we evaluate the contribution of Gaussian fluctuations to
the free energy of the mean field ordered state. The latter carries the
information embodied in the anisotropic spin interaction and therefore allows
to define preferred directions of the spin order with respect to the lattice.
We will not go beyond the simple evaluation of the contribution of
fluctuations, e.g., by incorporating the fluctuation contribution self-consistently.

For simplicity, we choose the parameters of the model such that the ground
state is ferromagnetic, i.e. we consider the Heisenberg interaction to be
ferromagnetic and allow the compass interaction to be both ferromagnetic and
antiferromagnetic. For any ferromagnetic
and weak antiferromagnetic compass interactions, the minima of the
fluctuational part of the free energy are attained if the spontaneous
magnetization vector points along one of the cubic axes. 

This paper is organized as follows. In section II we introduce the functional
integral representation of the partition function for the spin systems with
interactions described by the most general bilinear form of the super-exchange
Hamiltonian. The details of the method are outlined in the Appendix. In
Sec. III, we apply this framework to compute the angular dependence of the
fluctuational part of the free energy for the ferromagnetic Heisenberg-compass
model on the cubic lattice. Our results are presented and discussed in Section IV.

\section{Representation of the partition function}

\label{sec:partition function}

We consider a system of identical classical spins $\mathbf{S}$ on a lattice, interacting
in an anisotropic fashion as indicated in the introduction, defined by the
Hamiltonian
\begin{equation}
H=\frac{1}{2}\sum_{j,j^{\prime}}\,\sum_{\alpha\alpha^{\prime}}\,J_{j,j^{\prime
}}^{\alpha,\alpha^{\prime}}\,S_{j}^{\alpha}\,S_{j^{\prime}}^{\alpha^{\prime}},
\label{ham}%
\end{equation}
where $j,j^{\prime}$ label the lattice sites, $\alpha,\alpha^{\prime
}=x,y,z$ label the three components of the spin and $\mathbf{S}^2=1$. For the models with
compasslike anisotropic and Heisenberg isotropic interactions of spins, the
interaction is diagonal in spin space, $\alpha=\alpha^{\prime}$. The
$J_{j,j^{\prime}}^{\alpha,\alpha}$-matrix elements are different for the
$(j,j^{\prime})$-bonds along direction $\gamma$ with $\gamma=\alpha$ and
$\gamma\neq\alpha$. However, since our consideration is also valid for the
case when $\alpha\neq\alpha^{\prime}$, in the following, we will keep both indices.

We will be interested in the long-range ordered phases of the system. The mean
field approximation of the order parameter usually leads to a highly
degenerate manifold of states, e.g., a ferromagnetic state with spontaneous
magnetization pointing in any direction. This degeneracy is lifted by the
anisotropic components of the spin interaction, but only at the level of the
fluctuation contribution to the free energy (action) $S_{fl}$. In the
following, we outline a method allowing to calculate $S_{fl}$, which is based
on the Hubbard-Stratonovich transformation of the partition function for spin
systems described by the generic Hamiltonian (\ref{ham}). We present details and discuss justifications for this method in the Appendix.

The partition function of the system is given by the  integral over the Boltzmann weights of configurations
\begin{equation}
Z=\int[dS_{j}]\exp[-\beta\sum_{j\alpha,j^{\prime}\alpha^{\prime}}J_{jj^{\prime}%
}^{\alpha\alpha^{\prime}}S_{j}^{\alpha}S_{j^{\prime}}^{\alpha^{\prime}}]
\delta(\mathbf{S}_{j}^{2}-1),
\end{equation}
where $\beta=1/k_BT$ is the inverse temperature, $S_{j}^{\alpha}$ are the
components of the spin operator at site $j$.

It is useful to represent the
Hamiltonian in the basis of the eigenfunctions $\chi_{n;j,\alpha}$ of the spin
exchange matrix, defined by%
\[
\sum_{j^{\prime},\alpha^{\prime}}J_{jj^{\prime}}^{\alpha\alpha^{\prime}}%
\chi_{n;j^{\prime},\alpha^{\prime}}=\kappa_{n}\chi_{n;j,\alpha}\,.
\]
For spins on a periodic lattice these eigenstates are labeled by a wavevector
$\mathbf{q}$ (inside the first Brillouin zone) and index $\nu$,
characterizing three principle axes of the matrix ${\hat{J}}$. Thus
$|n\rangle=|\mathbf{q},\nu\rangle$ and the normalized eigenfunctions take the
form%
\[
\chi_{\mathbf{q},\nu;j,\alpha}=\frac{1}{\sqrt{N}}e^{i\mathbf{q\cdot R}_{j}%
}u_{\nu,\alpha}\,,
\]
where $N$ is the number of lattice sites, the  $u_{\nu,\alpha}$ are orthonormal
real-valued eigenvectors, i.e., $\sum_{\alpha}u_{\nu,\alpha}u_{\nu^{\prime
},\alpha}=\delta_{\nu\nu^{\prime}}$ and $\kappa_{\mathbf{q},\nu}$ are the
eigenvalues of the spin exchange interaction matrix.

We now define the normal amplitudes of the  spins as%
\[
S_{\mathbf{q},\nu}=\sum_{j,\alpha}\chi_{\mathbf{q},\nu;j,\alpha}S_{j}^{\alpha}%
\]
and express the Hamiltonian as%
\begin{eqnarray}
H=\sum_{\mathbf{q},\nu}\kappa_{\mathbf{q},\nu}S_{\mathbf{q},\nu}^{\ast
}S_{\mathbf{q},\nu}\,,
\end{eqnarray}
where $S_{\mathbf{q},\nu}^{\ast}=S_{-\mathbf{q},\nu}$.
 Commutation of classical spins allows us to employ a Hubbard-Stratonovich transformation in terms of
classical fields $\varphi_{\mathbf{q,}\nu}$ in order to represent the interaction
operator as a Zeeman energy operator of spins in a spatially varying
magnetic field. As a result, one finds the following representation of the
partition function:
\begin{eqnarray}
&&Z=\int[d\varphi]\\\nonumber
&&\exp\Bigl(-\beta\bigl[\sum_{\mathbf{q,}\nu}|\kappa_{\mathbf{q,}\nu}%
|^{-1}\varphi_{\mathbf{q,}\nu}^{\ast}\varphi_{\mathbf{q,}\nu}-{\mathcal{S}}%
_{loc}(\{\varphi_{\mathbf{q,}\nu}^{\ast},\varphi_{\mathbf{q,}\nu}\})\bigr]\Bigr),
\end{eqnarray}
where the integration volume element is given by
\[
\lbrack d\varphi]=\Pi_{\mathbf{q,}\nu}\frac{i\beta{d\varphi_{\mathbf{q,}\nu}^{\ast}%
}d\varphi_{\mathbf{q,}\nu}}{2\pi|\kappa_{\mathbf{q,}\nu}|}\,.
\]
The contribution  to the action in the case of classical
spins\ is given by%
\begin{equation}
S_{loc}(\{\varphi_{\mathbf{q,}\nu}^{\ast},\varphi_{\mathbf{q,}\nu}%
\})=\beta^{-1}\sum_{j}\ln\bigl[\sinh(2\beta{\varphi}_{j})/2\beta
{\varphi}_{j}\bigr],
\end{equation}
where ${\varphi}_{j}^{2}=({\varphi}_{j}^{x})^{2}%
+({\varphi}_{j}^{y})^{2}+({\varphi}_{j}^{z})^{2}$ , with
${\varphi}_{j}^{\alpha}\equiv\sum_{\mathbf{q,}\nu}s(\kappa_{\mathbf{q,}\nu
})\varphi_{\mathbf{q,}\nu}\chi_{\mathbf{q,}\nu;j,\alpha}^{\ast}$ and
$s(\kappa_{\mathbf{q,}\nu})=1$ for $\kappa_{\mathbf{q,}\nu}<0$  and $s(\kappa_{\mathbf{q,}\nu})=i$ for $\kappa_{\mathbf{q,}\nu}>0$. The
Hubbard-Stratonovich identity used to derive the above functional integral is
different for eigenmodes $\varphi_{\mathbf{q,}\nu}$ with positive or negative
eigenvalue $\kappa_{\mathbf{q,}\nu}$, leading to the appearance of a
complex-valued ${\varphi}_{j}$.
  The details of evaluating ${\mathcal{S}}_{loc}(\{\varphi_{\mathbf{q,}\nu}^{\ast},\varphi_{\mathbf{q,}\nu}\})$ can be found in the Appendix.

\section{Application to the cubic lattice}

\subsection{Isotropic Heisenberg interaction}

In order to demonstrate how to perform the evaluation of the above
representation of the partition function, we consider first the isotropic
ferromagnetic Heisenberg model with nearest neighbor interactions on the cubic
lattice. In this case, the Hamiltonian (\ref{ham}) reads
\begin{align}
H=J\sum_{\langle j;j^{\prime}\rangle}\sum_{\alpha}S_{j}^{\alpha}S_{j^{\prime}%
}^{\alpha},
\end{align}
where the lattice summation is over nearest neighbors $\langle j,j^{\prime
}\rangle-$bonds and $J<0$. For the isotropic exchange interaction, the eigenvalues, $\kappa
_{\mathbf{q,}\nu}=J\sum_{\alpha}\cos q_{\alpha}$, are independent of $\nu$,
$\kappa_{\mathbf{q,}\nu}=\kappa_{\mathbf{q}}$, and hence are degenerate.

A uniform ferromagnetic mean-field solution is
found by solving the saddle point equation
\begin{align}
\label{MF}
&\frac{\partial}{\partial\varphi_{{\small MF}}}{\mathcal{S}}
 =-\frac{\partial}{\partial\varphi_{{\small MF}}}N
\Bigl[|\kappa_{\mathbf{q=0}}|^{-1}(\varphi_{MF})^{2}\\\nonumber&
-\beta^{-1}\ln[\sinh(2\beta\varphi_{MF})/2\beta\varphi_{MF}]\Bigr]=0
,\nonumber
\end{align}
where we used $\varphi_{\mathbf{q},\nu}^{{\small MF}}=\sqrt{N}\varphi_{{\small MF}%
}\delta_{\mathbf{q},0}m_{0,\nu}$, ${\varphi}_{j}=\varphi_{{\small MF}}$,
 $m_{0,\nu}$ for the components of the unit vector along the magnetization
in the reference frame defined by the principal axes of the interaction matrix
(which are the cubic axes in this case), and $N$ is the number of lattice
sites.  The solution of  Eq.(\ref{MF})
gives us a non-linear  equation for the mean-field parameter:
\begin{equation}\label{eq:MF}
2|\kappa_{\mathbf{q=0}}
|^{-1}\varphi_{ MF}-2\coth(2\beta\varphi_{MF})+\frac{1}{\beta\varphi_{MF}}=0
\end{equation}
We solve this equation numerically  at each temperature and get $\varphi_{{\small MF}} (T)$. 
Linearizing Eq. (\ref{eq:MF}) near the transition,
we find the transition temperature $T_{c}=\beta_{c}^{-1}%
=2|\kappa_{\mathbf{q=0}}|/3$. We note in passing that a different length of the 
classical spin vector $|\mathbf{S}|=S_0$ may be simply scaled back to the unit length by 
changing the temperature as $T'=S_{0}^2 T$. Choosing $S_{0}^2=3/4$ appropriate 
for quantum spin $S=1/2$, we find the renormalized transition temperature $T'_{c}=|\kappa_{\mathbf{q=0}}|/2$ , which agrees with the quantum mean-field transition temperature.

The fluctuation contribution is obtained by expanding the action in the
fluctuation field $\delta\varphi_{\mathbf{q},\nu}=\varphi_{\mathbf{q},\nu}%
-\varphi_{\mathbf{q},\nu}^{{\small MF}}$ about the mean field solution to the lowest
order:
\begin{eqnarray}
{\mathcal{S}}  & =&{\mathcal{S}}_{0}+{\mathcal{S}}_{\mathrm{fl}}\\\nonumber
{\mathcal{S}}_{0}  &  =&
N|\kappa_{\mathbf{q=0}}|^{-1}
\varphi_{\small MF}^2
-NT\ln[\sinh(2\beta{\varphi}_{MF})/2\beta
{\varphi}_{MF}].
\end{eqnarray}

For Gaussian fluctuations, the fluctuation part of the free energy, or
equivalently the action, ${\mathcal{S}}_{\mathrm{fl}}$, is a bilinear function
of $\delta\varphi_{\mathbf{q,}\nu}$.  
It is given by
\begin{equation}
{\mathcal{S}}_{\mathrm{fl}}\{\delta\varphi_{\mathbf{q,}\nu}\}=\sum_{\mathbf{q}%
;\nu,\nu^{\prime}}A_{\mathbf{q,}\nu\nu^{\prime}}\delta\varphi_{\mathbf{q,}\nu
}^{\ast}\delta\varphi_{\mathbf{q,}\nu^{\prime}},\label{fluct-free-energy}%
\end{equation}
where we defined matrix elements of $A_{\mathbf{q},\nu\nu^{\prime}}$ describing the
weight of the Gaussian fluctuations of wavevector $\mathbf{q}$ and polarization
$\nu$ as
\begin{eqnarray}\label{fluct-matrix1}
&&A_{\mathbf{q},\nu\nu^{\prime}} =|\kappa_{\mathbf{q},\nu}|^{-1}\delta_{\nu
,\nu^{\prime}} 
 -\frac{2}{3}\Bigl[\beta_{c}(\delta_{\nu,\nu^{\prime}}-m_{0,\nu
}m_{0,\nu^{\prime}})+
\nonumber \\&&
3\beta rm_{0,\nu}m_{0,\nu^{\prime}}]s(\kappa
_{\mathbf{q},\nu})s(\kappa_{\mathbf{q},\nu^{\prime}})\Bigr]
\end{eqnarray}
 Here, for shortness we introduced  $r=1/(2\beta \varphi_{MF})^{2}-1/\sinh^{2}(2\beta\varphi_{MF})$.

\begin{figure}[ptb]
\label{fig1} \includegraphics[width=0.65\columnwidth]{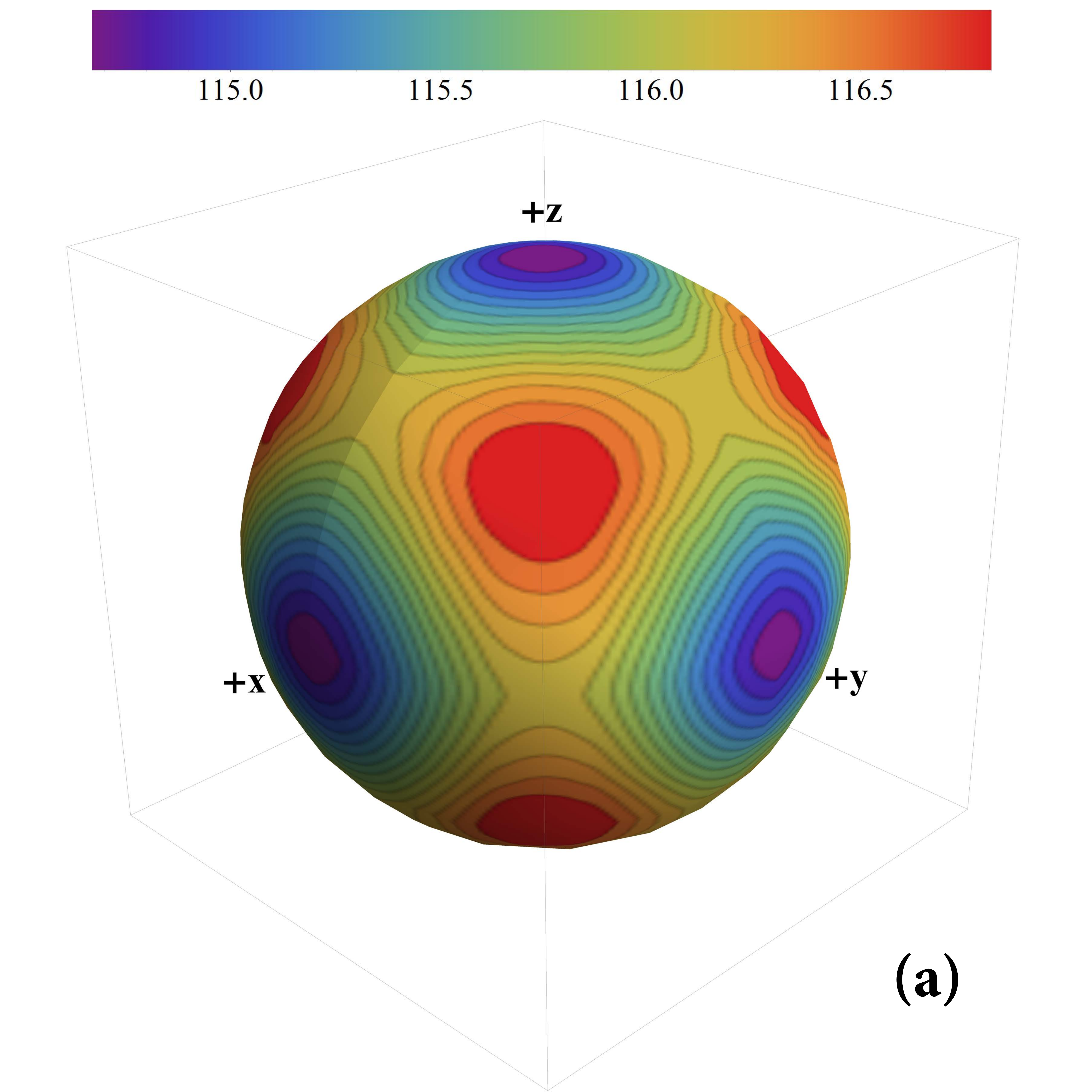}
\caption{(Colors online)
The magnitude of the action ${\mathcal{S}}_{\mathrm{fl}}(\theta,\phi)$ defined
by Eq. (\ref{action1}) is plotted on the surface of the unit sphere. The
preferred directions of the magnetization, corresponding to the minima of the
free energy, are shown by deep blue color. The energy scale is shown in units
of $J$. $J=-1$ and $K=0.75$: the preferred directions of the magnetization
are along the cubic axes.}%
\end{figure}

In the limit of small $\mathbf{q}$, it is instructive to separate the fluctuations into longitudinal (along
$\mathbf{m}_{0}$) and transverse (perpendicular to $\mathbf{m}_{0}$)
components, $\delta\varphi_{\bm q}^{\mathrm{l}}=\mathbf{m}_{0}\cdot\delta
{\bm\varphi}_{\bm q}$ and $\delta{\bm\varphi}_{\bm q}^{\mathrm{tr}}=\sum_{\mu
=1,2}\mathbf{m}_{\mu}\varphi_{{\bm q},\mu}^{\mathrm{tr}}$, respectively. We
defined $\delta\varphi_{{\bm q},\mu}^{\mathrm{tr}}=\mathbf{m}_{\mu}\cdot
\delta{\bm\varphi}_{\bm q}$, with $\mathbf{m}_{1}=(\mathbf{m}_{0}\mathbf{\times
z)/|\sin}\theta\mathbf{|}$ and $\mathbf{m}_{2}=\mathbf{m}_{1}\times
\mathbf{m}_{0}$, where $\cos\theta=\mathbf{m}_0\mathbf{\cdot z}$. 
Despite the complex nature of fluctuational fields, their separation into transverse and longitudinal modes is possible in the limit of small $\mathbf{q}$,  because
the interaction eigenvalues $\kappa_{\mathbf{q},\nu}<0$ and thus, 
$s(\kappa_{\mathbf{q},\nu})=1$ in this region of the BZ for any polarization component $\nu$.
Then, the longitudinal fluctuations
contribute to the free energy as
\begin{equation}
{\mathcal{S}}_{\mathrm{fl,l}}=\sum_{\bm q}
\left[ |\kappa_{\bm q}|^{-1}-2\beta r\right]
({\bm m}_{0}\mathbf{\cdot}\delta{\bm\varphi}_{\bm q}^{\mathrm{l}%
})(\mathbf{m}_{0}\mathbf{\cdot}\delta{\bm\varphi}_{-{\bm q}}^{\mathrm{l}}).
\end{equation}
 The transverse fluctuations
are gapless in agreement with Goldstone's theorem:%
\begin{equation}
{\mathcal{S}}_{\mathrm{fl,tr}}=\sum_{\bm q,\kappa_{\mathbf{q}}<0}%
\left[|\kappa_{\mathbf{q}}|^{-1}-\frac{2}{3}\beta_{c}\right](\delta{\bm\varphi}_{\mathbf{q}%
}^{\mathrm{tr}}\mathbf{\cdot}\delta{\bm\varphi}_{-\mathbf{q}}^{\mathrm{tr}})
\end{equation}
since $\lim_{\mathbf{q\rightarrow0}}\left[|\kappa_{\mathbf{q}}|^{-1}-\frac{2}%
{3}\beta_{c}\right]=0$.

\subsection{Fluctuations due to anisotropic compass interactions}

Next, in addition to the isotropic Heisenberg term, let us take into
consideration an anisotropic compass interaction, $K$. The constraint that the
ferromagnetic mean field solution remains stable is satisfied for all negative
(ferromagnetic) values of $K$ and for positive values $K<|J|$.

In the presence of the anisotropic compass interaction, the model (\ref{ham})
reads
\begin{align}
H=\sum_{j;j^{\prime}}\sum_{\alpha}J_{jj^{\prime}}^{\alpha}S_{j}^{\alpha
}S_{j^{\prime}}^{\alpha},
 \label{hamaniso}%
\end{align}
where the exchange interaction is given by
\begin{align}
J_{jj^{\prime}}^{\alpha}=\frac{1}{2}\delta_{j^{\prime},j+\tau}[J+K\delta
_{\alpha,|\tau|}]
\end{align}
The index $\tau=\pm x,\pm y,\pm z$ labels nearest neighbor sites, where
$|\tau|=x,y,z$\ specifies a direction in spin space ($x$ for bonds along the $x$-direction, etc.). The eigenvalues of the operator
$J_{jj^{\prime}}^{\alpha\alpha^{\prime}}$ defined in the previous section are
given by
\begin{align}
\kappa_{\mathbf{q},\nu}=\sum_{\alpha}(J+K\delta_{\alpha,\nu})\cos q_{\alpha}.%
\end{align}
The eigenvectors $\mathbf{u}_{\nu}$\ are again along the three cubic axes,
such that the components are $u_{\nu,\alpha}=\delta_{\nu,\alpha}$.\ This time
the three eigenvalues for given $\mathbf{q}$ are not degenerate (except in the
limit $\mathbf{q}\rightarrow0$) and the fluctuation contribution to the free
energy will therefore depend on the orientation of the spontaneous
magnetization.\ We may again use the representation of the partition function
$Z$ as a functional integral over the Fourier
components $\varphi_{\mathbf{q},\nu}$ of the auxilliary field.

Provided $J<0$ and $K<|J|$, the mean-field solution $\varphi_{MF}$ is given as
before by solving the transcendental equation (\ref{eq:MF}) numerically. The fluctuation contribution to the
free energy is obtained by expanding the action in the fluctuation field about
the mean field solution to lowest order. We get
\begin{equation}
Z=C\exp(-\beta{\mathcal{S}}_{0})\int[d\delta\varphi]\exp(-\beta{\mathcal{S}%
}_{\mathrm{fl}}\{\delta\varphi_{\mathbf{q,}\nu}\}),\label{sflaniso}%
\end{equation}
where the fluctuation part of the action is given by
Eqs.(\ref{fluct-free-energy}) and (\ref{fluct-matrix1}) In the following,
we show that by comparison to the isotropic model, Eq. (\ref{sflaniso})
manifestly breaks rotational invariance, which results in a selection of
preferred directions of the order parameter, which minimize the free energy.

The $3\times3$-matrix $A_{\mathbf{q,}\nu\nu^{\prime}}$ may be diagonalized and
has eigenvalues $\lambda_{\gamma,{\bm q}}$ and eigenvectors ${\bm v}%
_{\gamma,{\bm q}}$, $\gamma=0,1,2$. This allows us to express $\sum_{\nu
\nu^{\prime}}A_{\mathbf{q,}\nu\nu^{\prime}}\delta\varphi_{\mathbf{q,}\nu}^{\ast
}\delta\varphi_{\mathbf{q,}\nu^{\prime}}=\sum_{\gamma}\lambda_{\gamma,{\bm q}%
}\delta\varphi_{{\bm q},\gamma}\delta\varphi_{-{\bm q},\gamma}$, where $\delta
\varphi_{{\bm q},\gamma}={\bm v}_{\gamma,{\bm q}}\cdot\delta\boldsymbol{\varphi
}_{\bm q}$. The integration over the fluctuation amplitudes may now be
performed and gives
\begin{equation}
S_{\mathrm{fl}}=\beta^{-1}\frac{1}{2}\sum_{\mathbf{q}}\ln
|\lambda_{0,{\bm q}}\lambda_{1,{\bm q}}\lambda_{2,{\bm q}}|,\label{action1}%
\end{equation}
where we chose $s(\kappa_{\mathbf{q},\nu})=\pm i$ for $\kappa_{\mathbf{q},\nu}>0%
$, following the procedure described at the end of the Appendix.
Alternatively, we may use that $|\lambda_{0,{\bm q}}\lambda_{1,{\bm q}}%
\lambda_{2,{\bm q}}|=|\det\{A_{\mathbf{q,}\nu\nu^{\prime}}\}|$, saving the
trouble of having to determine the eigenstates of $A_{\mathbf{q,}\nu
\nu^{\prime}}$.

Let us now derive the explicit expression for the fluctuation contribution for
an arbitrary orientation of $\mathbf{m}_{0}=(\sin\theta\cos\phi,\sin\theta
\sin\phi,\cos\theta).$ Inserting this into the definition of
 $A_{\mathbf{q,}\nu\nu^{\prime}}$ given by Eq.(\ref{fluct-matrix1}), we find its elements to be
\begin{equation}%
\begin{array}
[c]{l}%
A_{{\bm q},00}=|\kappa_{{\bm q},x}|^{-1}-\frac{2}{3} s(\kappa_{{\bm q},x})s(\kappa_{{\bm q},x})
(\beta_c(1-s_{\theta}^{2}c_{\phi}^{2})+3\beta r s_{\theta}^{2}c_{\phi}^{2})
\\
A_{{\bm q},01}=-\frac{2}{3} s(\kappa_{{\bm q},x})s(\kappa_{{\bm q},y})(3\beta r-\beta_c)
c_{\phi}s_{\phi}s_{\theta}^2\\
A_{{\bm q},10}=A_{{\bm q},01}\\
A_{{\bm q},02}=-\frac{2}{3} s(\kappa_{{\bm q},x})s(\kappa_{{\bm q},z})(3\beta r-\beta_c)
c_{\phi}c_{\theta}s_{\theta}\\
A_{{\bm q},20}=A_{{\bm q},02}\\
A_{{\bm q},11}=|\kappa_{{\bm q},y}|^{-1}
-\frac{2}{3} s(\kappa_{{\bm q},y})s(\kappa_{{\bm q},y})
(\beta_c(1-s_{\theta}^{2}s_{\phi}^{2})+3\beta r s_{\theta}^{2}s_{\phi}^{2})\\
A_{{\bm q},12}=-\frac{2}{3} s(\kappa_{{\bm q},y})s(\kappa_{{\bm q},z})(3\beta r-\beta_c)
s_{\phi}c_{\theta}s_{\theta}\\
A_{{\bm q},21}=A_{{\bm q},12}\\
A_{{\bm q},22}=|\kappa_{{\bm q},z}|^{-1}
-\frac{2}{3} s(\kappa_{{\bm q},z})s(\kappa_{{\bm q},z})
(\beta_c s_{\theta}^{2}+3\beta r c_{\theta}^{2}),
\end{array}
\label{cubicmatrix}%
\end{equation}
where, to shorten notations, we denote $\sin\theta(\phi)\equiv s_{\theta
(\phi)}$ and $\cos\theta(\phi)\equiv c_{\theta(\phi)}$. The interactions are defined as $\kappa_{{\bm q},x}%
^{-1}=1/\left[(J+K)\cos q_{x}+J\cos q_{y}+J\cos q_{z}\right]$, $\kappa_{{\bm q},y}%
^{-1}=1/\left[(J+K)\cos q_{y}+J\cos q_{x}+J\cos q_{z}\right]$ and $\kappa_{{\bm q}%
,z}^{-1}=1/\left[(J+K)\cos q_{z}+J\cos q_{x}+J\cos q_{y}\right]$. We see that the matrix
$A_{{\bm q},\nu\nu^{\prime}}$ has a rather complex structure as a function of
${\bm q}$ and angles $\theta$ and $\phi$. This gives rise to a complex
behavior of the eigenvalues $\lambda_{0,{\bm q}}$, $\lambda_{1,{\bm q}}$ and
$\lambda_{2,{\bm q}}$.

\section{Results and discussions}

We now present the results obtained for ${\mathcal{S}}_{\mathrm{fl}}%
(\theta,\phi)$ by performing numerical integration in Eq.(\ref{action1}). The
angular dependence of ${\mathcal{S}}_{\mathrm{fl}}(\theta,\phi)$ is presented in Figs.1, where the magnitude of
$S_{\mathrm{fl}}(\theta,\phi)$ as a function of orientation of the spontaneous
magnetization is shown as a color-coded plot on the unit sphere. The
calculations in Figs.1 are performed at temperature
$\beta=\beta_{c}+1$ and assuming $J=-1$. We see that ${\mathcal{S}%
}_{\mathrm{fl}}(\theta,\phi)$ has a non-trivial dependence on the direction of
the order parameter defined by angles $\theta$ and $\phi$. This peculiar angular dependence
of ${\mathcal{S}}_{\mathrm{fl}}(\theta,\phi)$ is inherited from non-trivial
angular dependencies of $\lambda_{0,{\bm q}}$, $\lambda_{1,{\bm q}}$ and
$\lambda_{2,{\bm q}}$.

In Fig.1, we present the profile of ${\mathcal{S}}_{\mathrm{fl}}%
(\theta,\phi)$ computed for $K=0.75$. We can see that ${\mathcal{S}%
}_{\mathrm{fl}} (\theta,\phi)$ is minimized when the magnetization is directed
along one of the cubic axes. We note that the cubic directions are also
selected for other values of the compass interactions, both antiferromagnetic as well as ferromagnetic, where the ferromagnetic state is the mean field solution ($K<|J|$). 

\section{Conclusion}

The magnetic properties of heavy transition metal oxides such as iridates and
others are emerging as a new fascinating field offering opportunities to
realize strongly frustrated quantum spin systems in the laboratory. In these
systems, the combination of multiband electronic structure and strong Coulomb
and Hund's couplings with strong spin-orbit interaction can give rise to
extremely anisotropic spin exchange interactions of the compass type. Mean
field solutions of these models are often untouched by the anisotropies of the
model and show the full isotropy of pure Heisenberg models, in contrast with
experimental observations. In this paper, we addressed the question how the
system selects special preferred directions of the mean field order parameter
vector. We restricted ourselves to the case of a ferromagnetic order
parameter, but an analogous question exists for antiferromagnetic or more
complicated ordered structures. We find that the high degeneracy of the
ferromagnetic mean-field solution is lifted by the free energy contribution
from thermal fluctuations. We calculated the fluctuation
contribution for a Heisenberg-compass model of classical spins on a three
dimensional cubic lattice with nearest neighbor interactions - an isotropic
Heisenberg coupling $J<0$ (which we take as the energy unit), and a compass
coupling $K$. The ferromagnetic state is found if $K<|J|$. Rather than
exploring the full phase diagram, we focused on one typical temperature
$T=T_{c}/(1+T_{c})$, where $T_{c}$ is the mean-field transition temperature.
For values of $K<1$, the system is
found to choose preferred directions of the spontaneous magnetization along
one of the cubic axes. In the temperature regime 
considered here, we expect the classical approximation to be valid. 
A generalization to quantum spin systems of the approach presented here 
is in preparation.

\textit{Acknowledgements.} We thank Michel Gingras, George Jackeli, Yoshi
Kamiya, Alberto Hinojosa-Alvarado and Ioannis Rousochatzakis for useful discussions. N.P. and Y.S.
acknowledge the support from NSF Grants DMR-1005932 and DMR-1511768. P.W.
thanks the Department of Physics at the University of Wisconsin-Madison for
hospitality during several stays as a visiting professor. P.W. also
acknowledges partial support by an ICAM senior fellowship and through the DFG
research unit "Quantum phase transitions". N.P. acknowledges the hospitality
of KITP and partial support by the National Science Foundation under Grant No.
NSF PHY11-25915.

\appendix

\section{Hubbard-Stratonovich transformation of the partition function for
spin systems}

\subsection{General formulation}

The Hubbard-Stratonovich (H-S) transformation is based on the mathematical identitiy%

\begin{equation}
\exp[-ax^{2}]=\frac{1}{\sqrt{\pi|a|}}\int dy\exp\left[-\frac{y^{2}}{|a|}+2s(a)xy)\right],
\end{equation}
where we defined
\begin{equation}
s(a)={\Big \{}%
\begin{array}
[c]{cc}%
1\,, & \mathrm{if}\,a<0\\
\imath\,, & \mathrm{if}\,a>0.
\end{array}
\end{equation}
For $a>0$ we may as well use $s(a)=-i$. We will later make use of this
ambiguity when we evaluate the $y$-integrals approximately, which may lead to
imaginary-valued contributions.\ 

In the above H-S-transformation, $x$ may be a number or an operator. In the
case it is an operator, we use the eigenfunctions $|n\rangle$\ of
$\widehat{x}$ defined by
\[
\widehat{x}|n\rangle=x_{n}|n\rangle
\]
to prove that

\begin{eqnarray}
&& \exp
\left[-a\widehat{x}^{2}\right]|n\rangle=
\exp\left[-ax_{n}^{2}\right] |n\rangle\\
&&  =\frac{1}{\sqrt{\pi|a|}}\int dy
\exp\left[-\frac{y^{2}}{|a|}+2s(a)x_{n}%
y)
\right] |n\rangle \nonumber \\
&&  =\frac{1}{\sqrt{\pi |a|}}\int dy
\exp\left[-\frac{y^{2}}{|a|}+2s(a)\widehat{x}%
y)\right] |n\rangle .
\nonumber
\end{eqnarray}

This identity also works for complex (non-Hermitian) $x$ and $y$:
\begin{equation}
\exp[-a\widehat{x}^{\dag}\widehat{x}]=\frac{i}{2\pi |a|}\int dy^{\ast}%
dy\exp\left[-\frac{y^{\ast}y}{|a|}+s(a)(\widehat{x}^{\dag}y+H.c.)\right]\nonumber
\end{equation}

We now turn to the case of the partition function of a spin system with
generic interaction Hamiltonian (1). In order to use the mathematical
identities we need to represent the Hamiltonian (1) in terms of normal
coordinates. To this end we define the normalized eigenstates of the exchange
interaction operator
\begin{equation}
\sum_{j^{\prime},\alpha^{\prime}}J_{jj^{\prime}}^{\alpha\alpha^{\prime}}%
\chi_{n;j^{\prime},\alpha^{\prime}}=\kappa_{n}\chi_{n;j,\alpha},%
\end{equation}
in terms of which we have%
\begin{equation}
J_{jj^{\prime}}^{\alpha\alpha^{\prime}}=\sum_{n}\kappa_{n}\chi_{n;j,\alpha
}^{\ast}\chi_{n;j^{\prime},\alpha^{\prime}},
\end{equation}
where  $\chi_{n;j^{\prime},\alpha^{\prime}}$ form a complete and
orthonormal set of eigenfunctions and thus obey
\begin{align}
&  \sum_{j,\alpha}\chi_{n;j,\alpha}^{\ast}\chi_{n^{\prime};j,\alpha}%
=\delta_{n,n^{\prime}},\\
&  \sum_{n}\chi_{n;j,\alpha}^{\ast}\chi_{n;j^{\prime},\alpha^{\prime}}%
=\delta_{j,j^{\prime}}\delta_{\alpha,\alpha^{\prime}}.\nonumber
\end{align}

For spins on a periodic lattice, the eigenstates $|n\rangle=|\mathbf{q},\nu\rangle$ are labeled by wavevector $\mathbf{q}$ and spin component $\nu$,
and the eigenfunctions take the form%
\begin{equation}
\chi_{\mathbf{q},\nu;j,\alpha}=\frac{1}{\sqrt{N}}e^{i\mathbf{q\cdot R}_{j}%
}u_{\mathbf{q}\nu}^{\alpha}%
\end{equation}
where $u_{\mathbf{q}\nu}^{\alpha}$ are normalized real valued eigenvectors,
i.e. $\sum_{\alpha}u_{\mathbf{q}\nu}^{\alpha}u_{\mathbf{q},\nu}^{\alpha}=1$,
and $\kappa_{\mathbf{q},\nu}$ are the eigenvalues of the spin exchange operator.
We now define the normal amplitudes of the spin operators as
\begin{equation}
S_{\mathbf{q},\nu}=\sum_{j,\alpha}\chi_{\mathbf{q},\nu;j,\alpha}S_{j}^{\alpha}%
\end{equation}
and express the Hamiltonian (1) as%
\begin{equation}
H=\sum_{\mathbf{q},\nu}\kappa_{\mathbf{q},\nu}S_{\mathbf{q},\nu}^{\ast
}S_{\mathbf{q},\nu},%
\end{equation}
where $S_{\mathbf{q},\nu}^{\ast}=S_{-\mathbf{q},\nu}$.

We seek to apply the above mathematical identities (A1)-(A3) to each normal
component separately. This requires the normal components of the spin
operators to commute with each other, which is certainly true for the classical spins.
 Then using the Hubbard-Stratonovich transformation one may express
the Boltzmann weight operator of each normal mode in terms of normal field
amplitudes $\varphi_{\mathbf{q,}\nu}$ as%
\begin{align}
&  \exp[-\beta\kappa_{\mathbf{q,}\nu}S_{\mathbf{q,}\nu}^{\ast}S_{\mathbf{q,}%
\nu}]=
  \frac{\imath\beta}{2\pi|\kappa_{\mathbf{q,}\nu}|}\int\int d\varphi
_{\mathbf{q,}\nu}^{\ast}d\varphi_{\mathbf{q,}\nu}\\
&  \exp\left[-\beta\{|\kappa_{\mathbf{q,}\nu}|^{-1}\varphi_{\mathbf{q,}\nu}^{\ast}%
\varphi_{\mathbf{q,}\nu}+s(\kappa_{\mathbf{q,}\nu})(S_{\mathbf{q,}\nu}^{\ast}%
\varphi_{\mathbf{q,}\nu}+H.c.)\}\right]\nonumber
\end{align}
The complete Boltzmann weight operator may be expressed, again using the
commutability of the normal mode operators, as
\begin{align}
&  \exp[-\beta\sum_{\mathbf{q},\nu}\kappa_{\mathbf{q},\nu}S_{\mathbf{q},\nu
}^{\ast}S_{\mathbf{q},\nu}]=\int[d\varphi]\\
&  \exp[-\beta\sum_{\mathbf{q,}\nu}\{|\kappa_{\mathbf{q,}\nu}|^{-1}%
\varphi_{\mathbf{q,}\nu}^{\ast}\varphi_{\mathbf{q,}\nu}+s(\kappa_{\mathbf{q,}\nu
})(S_{\mathbf{q,}\nu}^{\ast}\varphi_{\mathbf{q,}\nu}+h.c.)\}],\nonumber
\end{align}
where $\varphi_{\mathbf{q,}\nu}^{\ast}=\varphi_{-\mathbf{q,}\nu}$\ . The
integration volume element is given by
\[
\lbrack d\varphi]=\Pi_{\mathbf{q,}\nu}\frac{i\beta{d\varphi_{\mathbf{q,}\nu}^{\ast}%
}d\varphi_{\mathbf{q,}\nu}}{2\pi|\kappa_{\mathbf{q,}\nu}|}%
\]
 Next, we find that the partition function of an
interacting classical spin system on an infinite periodic lattice may be
expressed as
\begin{align}\label{ZZ}
&Z=\\
&  =C\int[d\varphi]\exp\left[-\beta\sum_{\mathbf{q,}\nu}|\kappa_{\mathbf{q,}\nu}%
|^{-1}\varphi_{\mathbf{q,}\nu}^{\ast}\varphi_{\mathbf{q,}\nu}-S_{loc}(\{\varphi
_{\mathbf{q,}\nu}\})\right],\nonumber
\end{align}
where $C$ is a constant.\ The contribution ${\mathcal{S}}_{loc}(\{\varphi
_{\mathbf{q,}\nu}\})$ to the action is given by%
\begin{equation}
{\mathcal{S}}_{loc}(\{\varphi_{\mathbf{q,}\nu}\})=\frac{1}{\beta}\sum_{j}%
\ln W_j 
\end{equation}
 and  $W_{j} $ is computed by taking into account the constraint of the unit length of classical spins, $\mathbf{S}_{j}^{2}=1$, 
and  integrating over all directions of spin at each lattice site:%
\begin{align}
W_{j}  &  =\int\frac{dS_{j}d\Omega_{j}}{2\pi}\exp\left[2\beta\sum_{\alpha
}{\varphi}_{j}^{\alpha}S_{j}^{\alpha}\right]\delta(\mathbf{S}_{j}^{2}-1)\nonumber\\
&  =\int\frac{d\Omega_{j}}{4\pi}\exp\left[2\beta\sum_{\alpha}{\varphi}%
_{j}^{\alpha}S_{j}^{\alpha}\right]\\\nonumber
&  =\frac{\sinh\,2\beta|{\varphi}_{j}|}{2\beta|{\varphi}_{j}|}.%
\end{align}
This gives
\begin{align}\label{SlocAP}
{\mathcal{S}}_{loc}(\{\varphi_{\mathbf{q,}\nu}\})=\frac{1}{\beta}
\sum_{j}%
\ln[\frac{\sinh2\beta|{\varphi}_{j}|}{2\beta|{\varphi}_{j}|}].%
\end{align}
Here we defined  the complex-valued three-component field ${\varphi}%
_{j}^{\alpha}$ at each lattice site $j$  as
\begin{align}
{\varphi}_{j}^{\alpha}  &  =\sum_{\mathbf{q,}\nu}s(\kappa
_{\mathbf{q,}\nu}){\mathcal{R}e}\{\varphi_{\mathbf{q,}\nu}^{\ast}\chi
_{\mathbf{q,}\nu;j,\alpha}\}\\\nonumber
&  =\sum_{\mathbf{q,}\nu}s(\kappa_{\mathbf{q,}\nu})\varphi_{\mathbf{q,}\nu}%
\chi_{\mathbf{q,}\nu;j,\alpha}^{\ast}\nonumber\\
&  ={\varphi}_{R,j}^{\alpha}+i{\varphi}_{I,j}^{\alpha}.\nonumber
\end{align}
Observing that $\kappa_{\mathbf{q,}\nu}=\kappa_{-\mathbf{q,}\nu}$, we get%
\begin{align}\label{IR}
{\varphi}_{R,j}^{\alpha}  &  =\operatorname{Re}\{{\varphi}%
_{j}^{\alpha}\}=\sum_{\mathbf{q,}\nu,\kappa_{\mathbf{q,}\nu}<0}\varphi
_{\mathbf{q,}\nu}\chi_{\mathbf{q,}\nu;j,\alpha}^{\ast}\\\nonumber
{\varphi}_{I,j}^{\alpha}  &  =\operatorname{Im}\{{\varphi}%
_{j}^{\alpha}\}=\sum_{\mathbf{q,}\nu,\kappa_{\mathbf{q,}\nu}>0}\varphi
_{\mathbf{q,}\nu}\chi_{\mathbf{q,}\nu;j,\alpha}^{\ast}.%
\end{align}
 The field amplitude  is determined by
\begin{eqnarray}
{\varphi}_{j}=\sqrt{({\boldsymbol\varphi}_{R,j}+i\,{\boldsymbol\varphi}_{I,j})^{2}},
\end{eqnarray}
 where $\boldsymbol{\varphi}_{R,j} =\left({\varphi}^x_{R,j}, \,{\varphi}^y_{R,j},\,{\varphi}^z_{R,j}\right)$ and $\boldsymbol{\varphi}_{I,j} =\left({\varphi}^x_{I,j}, \,{\varphi}^y_{I,j},\,{\varphi}^z_{I,j}\right)$. 

We now derive the contribution of Gaussian fluctuations to the free energy for
the ferromagnetic mean field state which we denote as $\boldsymbol{\varphi}_{MF}$.
To this end,  we expand ${\mathcal{S}}_{loc}(\{\varphi_{\mathbf{q,}\nu}\})$ (\ref{SlocAP})   in terms of  the fluctuation amplitudes and  separate  the mean-field and fluctuational contributions. 
First, we  expand the field amplitude
${\varphi}_{j}$ to bilinear order in the fluctuation amplitudes:%
\begin{align}
{\varphi}_{j} &  ={\varphi}_{MF}+\delta {\varphi}_{j} ,\\\nonumber
\delta {\varphi}_{j}  &  =\frac{1}{2\varphi_{MF}}[2{\boldsymbol\varphi}_{MF}\cdot(\delta{\boldsymbol\varphi}
_{R,j}\mathbf{+}i\delta{\boldsymbol\varphi}_{I,j})+\delta{\boldsymbol\varphi}_{R,j}^{2}\,-%
\delta{\boldsymbol\varphi}_{I,j}^{2}]\\\nonumber
&  -\frac{1}{2\varphi_{MF}^3}[{\boldsymbol\varphi}_{MF}\cdot(\delta{\boldsymbol\varphi}_{R,j}%
+i\,\delta{\boldsymbol\varphi}_{I,j})]^{2}.
\end{align}
   Using Eq. (\ref{IR}), we now obtain
 the expressions for $\delta \varphi_{j} $ and $\delta \varphi_{j}^{2}$ in terms of
\ $\varphi_{\mathbf{q,}\nu}^{\ast}$ and $\varphi_{\mathbf{q,}\nu}$, keeping quadratic (Gaussian) terms only:
\begin{eqnarray}
&& \sum_{j}\delta \varphi_{j}   =\frac{1}{2{\varphi}_{MF}}\sum_{\mathbf{q},\nu,\nu^{\prime}%
}\delta_{\nu,\nu^{\prime}}s(\kappa_{\mathbf{q},\nu})s(\kappa_{\mathbf{q},
\nu^{\prime}})\varphi_{\mathbf{q,}\nu}^{\ast}\varphi_{\mathbf{q,}\nu^{\prime}}%
\nonumber\\ &&
-\frac{1}{2{\varphi}_{MF}}\sum_{j}\delta \varphi_{j}^{2}\\\nonumber
&& \sum_{j}\delta \varphi_{j}^{2}   =\sum_{\mathbf{q},\nu,\nu^{\prime}}s(\kappa
_{\mathbf{q},\nu})s(\kappa_{\mathbf{q},\nu^{\prime}})m_{0,\nu}\varphi
_{\mathbf{q,}\nu}^{\ast}\varphi_{\mathbf{q,}\nu^{\prime}}m_{0,\nu^{\prime}}.%
\end{eqnarray}

Next, we expand Eq.  (\ref{SlocAP}) step by step as
\begin{eqnarray}
&&\sinh2\beta|{\varphi}_{j}|  =\sinh(2\beta({\varphi}_{MF}+\delta \varphi_j))\nonumber\\\nonumber
&&=\sinh(2\beta{\varphi}_{MF})[1+2(\beta\delta \varphi_j)^{2}]+\cosh(2\beta {\varphi}_{MF}
)2\beta\delta \varphi_j
\end{eqnarray}
 and further
\begin{eqnarray}
&&\ln\left[\sinh(2\beta{\varphi}_{j})/2\beta{\varphi}_{j}\right]  \nonumber\\
&&=\ln\left[
\sinh(2\beta({\varphi}_{MF}+\delta\varphi_j ))\right]-\ln\left[2\beta({\varphi}_{MF}+\delta \varphi_j)\right]\nonumber\\
\nonumber
&&  =\ln\left[\sinh(2\beta {\varphi}_{MF})/(2\beta {\varphi}_{MF})\right]
\\\nonumber
&&+\left[2\beta {\varphi}_{MF}\coth(2\beta
{\varphi}_{MF})-1\right]\frac{\delta \varphi_j}{{\varphi}_{MF}}\\\nonumber
 && +\frac{1}{2}\left[-\frac{(2\beta {\varphi}_{MF})^{2}}{\sinh^{2}(2\beta {\varphi}_{MF})}%
+1\right](\frac{\delta \varphi_j}{{\varphi}_{MF}})^{2}.%
\end{eqnarray}
\noindent
The fluctuation part of the local part of the free energy is then given by%
\begin{eqnarray}
&&-\beta^{-1}\delta\sum_{j}\ln[\sinh(2\beta{\varphi}%
_{j})/2\beta{\varphi}_{j}]=\nonumber\\
&&  =-\frac{4}{3}\beta_{c} {\varphi}_{MF}\sum_{j}\delta \varphi_{j}\\\nonumber&&
-\frac{1}{2\beta  {\varphi}_{MF}^{2}%
}[1-\frac{(2\beta {\varphi}_{MF})^{2}}{\sinh^{2}(2\beta  {\varphi}_{MF})}]\sum_{j}\delta
\varphi_{j}^{2},%
\end{eqnarray}
where  we have used that $2\beta {\varphi}_{MF}\coth(2\beta {\varphi}_{MF})-1=\frac{4}{3}\beta_{c}\beta {\varphi}_{MF}^{2}$ .
Substituting the expressions for $\delta \varphi_{j},\delta \varphi_{j}^{2}$ and defining $r=1/(2\beta {\varphi}_{MF})^{2}-1/\sinh^{2}(2\beta{\varphi}_{MF})$, we get  the fluctuation contribution to the free energy
\begin{equation}
{\mathcal{S}}_{\mathrm{fl}}\{\delta\varphi_{\mathbf{q,}\nu}\}=\sum_{\mathbf{q}%
;\nu,\nu^{\prime}}A_{\mathbf{q,}\nu\nu^{\prime}}\delta\varphi_{\mathbf{q,}\nu
}^{\ast}\delta\varphi_{\mathbf{q,}\nu^{\prime}},\label{fluct-free-energyAP}%
\end{equation}
where we defined matrices $A_{\mathbf{q,}\nu\nu^{\prime}}$ describing the
weight of Gaussian fluctuations of wavevector $\mathbf{q}$ and polarization
$\nu$ as
\begin{eqnarray}\label{fluct-matrix1AP}
&&A_{\mathbf{q},\nu\nu^{\prime}} =|\kappa_{\mathbf{q},\nu}|^{-1}\delta_{\nu
,\nu^{\prime}}\\\nonumber 
&& -\frac{2}{3}[\beta_{c}(\delta_{\nu,\nu^{\prime}}-m_{0,\nu
}m_{0,\nu^{\prime}})+3\beta rm_{0,\nu}m_{0,\nu^{\prime}}]s(\kappa
_{\mathbf{q},\nu})s(\kappa_{\mathbf{q},\nu^{\prime}})
\end{eqnarray}

The fluctuation matrix $A_{\mathbf{q},\nu\nu^{\prime}}$ will in general be
non-Hermitian, and its eigenvalues will be complex. We now use that
$A_{\mathbf{q},\nu\nu^{\prime}}$ is an even function of $\mathbf{q}$ and divide
$\mathbf{q}$-space into $q_{x}>0$ ($M_{>}$) and $q_{x}<0$ ($M_{<}$). 
Note that the choice of $q_x$ for dividing the BZ in half is arbitrary, and we could also do it with a help of $q_y$ and $q_z$. 
For modes $\varphi_{\mathbf{q},\nu}$ with $\mathbf{q\in}M_{>}$, we choose
$s(\kappa_{\mathbf{q},\nu})=+i$, whereas for modes with $\mathbf{q\in}M_{<}$ we choose
$s(\kappa_{\mathbf{q},\nu})=-i$, where $\kappa_{\mathbf{q},\nu}>0$ in both
cases. Then we have $A_{-\mathbf{q},\nu\nu^{\prime}}=A_{\mathbf{q},\nu
\nu^{\prime}}^{\ast}$ and \ as a result of the functional integration we will
get
\begin{align}
Z &  =Z_{MF}\int[d\delta\varphi]\exp\left[-\beta\sum_{\mathbf{q,}\nu,\nu^{\prime}}A_{\mathbf{q},%
\nu\nu^{\prime}}\delta\varphi_{\mathbf{q,}\nu}^{\ast}\delta\varphi_{\mathbf{q,}\nu^{\prime}%
}\right]\nonumber\\
&  =Z_{MF}\exp\left[-\frac{1}{2}\sum_{\mathbf{q\in}M_{>}}\ln(\det(A_{\mathbf{q},\nu\nu^{\prime}%
})\det(A_{\mathbf{q},\nu\nu^{\prime}}^{\ast}))\right]\nonumber\\
&  =Z_{MF}\exp\left[-\frac{1}{2}\sum_{\mathbf{q}}\ln|\det(A_{\mathbf{q},\nu\nu^{\prime}})|\right],
\end{align}
where
 \begin{eqnarray}Z_{MF}=\exp\left[ -\beta N |\kappa_{{\bm q}=0}|^{-1}\varphi_{MF}^2 \right]\left[\sinh(2\beta {\varphi}_{MF})/(2\beta {\varphi}_{MF})\right]^N.
 \nonumber\end{eqnarray}


\begin{thebibliography}{99}                                                                                               %


\bibitem {nussinov15}Z. Nussinov, J. van den Brink, Rev. Mod. Phys.
\textbf{87} 1 (2015).

\bibitem {kk82}K. I. Kugel and D. I. Khomskii, Sov. Phys. Usp. \textbf{25},
231 (1982).

\bibitem {nussinov04}Z. Nussinov, M. Biskup, L. Chayes, and J. v. d. Brink,
Europhys. Lett. \textbf{67}, 990 (2004).

\bibitem {nussinov05}Z. Nussinov and E. Fradkin, Phys. Rev. B \textbf{71},
195120 (2005).

\bibitem {cincio10}
L. Cincio, J. Dziarmaga, and A. M. Oles, Phys. Rev. B \textbf{82},
104416 (2010).

\bibitem {nasu12}
J. Nasu, S. Todo, and S. Ishihara Phys. Rev. B \textbf{85}, 205141 (2012).

\bibitem {oles13}W. Brzezicki and A. M. Oles.
Phys. Rev. B \textbf{87}, 214421 (2013).

\bibitem {jackeli09}G. Jackeli and G. Khaliullin, Phys. Rev. Lett.
\textbf{102}, 017205 (2009).

\bibitem {batista05}C. Batista and Z. Nussinov, Phys. Rev. B \textbf{72},
045137 (2005).

\bibitem {kitaev06}A. Kitaev, Ann. Phys. \textbf{321}, 2 (2006).

\bibitem {biskup05}M. Biskup, L. Chayes, Z. Nussinov, Comm. Math. Phys.
\textbf{255}, 253 (2005).

\bibitem {jackeli10}J. Chaloupka, G. Jackeli, and G. Khaliullin, Phys. Rev.
Lett. \textbf{105}, 027204 (2010).

\bibitem {chern10}G.-W. Chern, N. Perkins, and Z. Hao Phys. Rev. B
\textbf{81}, 125127 (2010).

\bibitem {oitmaa10}J. Oitmaa, C. J. Hamer, Phys. Rev. B \textbf{83}, 094437 (2010).

\bibitem {fabien10}F. Trousselet, A. j M. Oles, Peter Horsch, European
Physical Letters \textbf{91}, 40005 (2010).

\bibitem {khali2001}G. Khaliullin, Phys. Rev. B \textbf{64}, 212405 (2001).

\bibitem {wenzel10}S. Wenzel, W. Janke and A. Lauchli, Phys. Rev. E
\textbf{81}, 066702 (2010)

\bibitem {perkins14}N.  B. Perkins, Y. Sizyuk and P. W\"{o}lfle,
Phys. Rev. B \textbf{89}, 035143 (2014).

\bibitem {sizyuk14}Y. Sizyuk, C. Price, P. W\"{o}lfle, and N. B.
Perkins, Phys. Rev. B \textbf{90}, 155126 (2014).

\bibitem {plakida14}A. A. Vladimirov, D. Ihle, N. M. Plakida, arxiv: 1411.3920.

\bibitem {gerlach15}M. H. Gerlach and W. Janke, Phys. Rev. B.
\textbf{91}, 045119 (2015).
\end{thebibliography}
\end{document}